\documentclass[aps,groupedaddress,
amsmath,  twocolumn,prl]{revtex4}

\usepackage{graphicx}
\usepackage{amsfonts}
\usepackage{amssymb}
\usepackage{amsbsy}
\usepackage{amsmath}
\usepackage{latexsym}
\usepackage{natbib}
\usepackage{bm}
\usepackage{subfigure}
\usepackage{color}
\usepackage{hyperref}
\usepackage{lscape}

\newcommand*{\di}{\partial}

\def\be {\begin{equation}}
\def\ee  {\end{equation}}
\def\bea {\begin{eqnarray}}
\def\eea {\end{eqnarray}}

\renewcommand*{\k}{\hat{k}}

\newcommand*{\M}{M_{\star}}

\def\k{\mathbf{k}}
\def\x{\mathbf{x}}
\def\y{\mathbf{y}}

\begin{document}

\title{Generalized uncertainty principles and quantum field theory}

\author{Viqar Husain}

\author{Dawood Kothawala}
 
\author{Sanjeev S. Seahra}

\affiliation{ Department of Mathematics and Statistics, University
of New Brunswick, Fredericton, NB, Canada E3B 5A3} 

\date{\today}

\begin{abstract}
  
Quantum mechanics with a generalized uncertainty principle arises through a representation of the 
commutator $[\hat{x}, \hat{p}] = i f(\hat{p})$. We apply this deformed  quantization to free scalar field theory for
$f_\pm =1\pm \beta p^2$. The resulting quantum field theories  have a rich fine scale structure.  For small wavelength
modes, the Green's function for $f_+$ exhibits a remarkable transition from Lorentz to Galilean invariance,
whereas for $f_-$ such modes effectively do not  propagate. For both cases  Lorentz  invariance is recovered  at  long wavelengths.

\end{abstract}

\maketitle

\paragraph{\bf Introduction}

One of the most important problems in fundamental physics is an understanding of  the high energy behaviour of quantum fields. This question is intimately connected with  the structure of spacetime at short distances, because the background mathematical structure that underlies quantum field theory (QFT), namely a manifold with a metric, may come into question in this regime. A part of the problem is that the spacetime metric forms a reference not only for defining the particle concept, but also for the Hilbert space inner product; if the metric is subject to quantum fluctuations then its use in an inner product becomes an issue.

There are many approaches that have been deployed to probe such questions, including string theory, non-commutative geometry, loop quantum gravity and causal sets.  Some of these suggest that the fundamental commutator $[\hat{x},\hat{p}]=i$ of quantum mechanics is modified at high energies.  For example, the particular modification
\begin{equation}\label{eq:QM commutator}
[\hat{x},\hat{p}] = if(\beta^{1/2} \hat{p}), \quad f(\beta^{1/2} \hat{p}) = 1 + \beta \hat{p}^{2},
\end{equation}
with (dimensionful) constant $\beta > 0$ has been studied for a number of systems, including the simple harmonic oscillator \cite{KMM}. It has also been used in the cosmological context to compute modifications to the spectrum of fluctuations in cosmology \cite{hassan}.  Recent experiments have attempted to put  constraints on $\beta$ \cite{GUPexp:2011}. However no direct application to QFT has  so far been studied.

 In this paper, we  apply the commutator algebra (\ref{eq:QM commutator}) to QFT in  flat spacetime for both generic and specific choices of the
 function $f$.   Our approach involves applying a 3-dimensional spatial Fourier transform to the classical phase space variables, and then enforcing the deformed commutator in $\k$-space. This approach was used for polymer quantization of the scalar field in \cite{HHS-prop} following work on a Fock-like quantization in \cite{HK-fock}.  

\paragraph{\bf Quantized scalar field}

We start with the Hamiltonian of a free scalar field in Minkowski space time:
\begin{eqnarray}\label{eq:standard H}
H_{\phi} &=& \int d^{3} x \, \frac{1}{2} \left[ \pi^{2} + (\mathbf{\nabla}\phi)^{2} \right],
\end{eqnarray}
where $(\phi, \pi)$ satisfy $\{\phi(t,\x),\pi(t,\y)\} = \delta^{(3)}(\x-\y)$. The Fourier modes are 
 \begin{align}
\phi(t,\x) &=  \frac{1}{\sqrt{V}}\sum_{\k}{\phi}_{\k}(t) e^{i {\k}\cdot{\x}},
\end{align}
with a similar expansion for $\pi(t,\x)$; $V = \int d^{3}x$ is a fiducial volume for box normalization.
After  a suitable redefinition of independent modes to enforce that $\phi$ is real, the Hamiltonian becomes
\begin{equation}\label{eq:Fourier H}
H_{\phi} =  \sum_{\k} H_{\k} = \sum_{\k} \frac{1}{2} \left[ \pi_{\k}^2 +
k^{2} \phi_{\k}^2 \right], \quad k = |\k|,
\end{equation}
where the $\k$-space canonical variables satisfy the Poisson bracket $\{\phi_\k, \pi_{\k '}\} = \delta_{\k,\k'}$. The structure of the Hamiltonian is that of a collection of decoupled simple harmonic oscillators labelled by $\k$, therefore the obvious Hilbert space for constructing the quantum theory is  a tensor product
${\cal H} =\otimes_\k {\cal H}_\k$. 

We quantize field theory by  representing the  modified commutator on the $\k$-space canonical variables:
\begin{equation}\label{eq:modified commutator}
	[\hat{\phi}_{\k} ,\hat{\pi}_{\k} ] = i  f( \hat{\pi}_{\k}/M_{\star}^{1/2}),
\end{equation}
where $f$ is a dimensionless function and $\M$ is an energy scale.   In the momentum space representation  with 
$\psi(\pi_\k) \in  {\cal H}_\k = L^2(I,f^{-1}d\pi_{\k})$, the modified commutator is realized by the operator definitions 
\begin{subequations}
\begin{align}
\hat{\phi}_\k  \psi(\pi_\k) &=  i f (\pi_{\k}/M_{\star}^{1/2}) {\di_{\pi_{\k}}} \psi(\pi_{\k})\\
\hat{\pi}_\k \psi(\pi_\k) &= \pi_\k  \psi(\pi_\k).
\end{align}
\end{subequations}
The interval $I$ must be selected such that $f \ge 0$ for all $\pi_{\k} \in I$.  Although the function $f$  may be arbitrary up to the action of operators still giving  $L^2$ functions,  we  impose the additional condition
that $f(0) = 1$  to recover the standard commutator for small momenta $\pi_{\k} \ll M_{\star}^{1/2}$. This enforces the requirement that the  effects of deformation are confined to short wavelengths, as we shall see in the following. 

In this representation, the energy eigenvalue equation $H_\k \psi = E^{\k}_n \psi$ reads 
\begin{subequations}\label{eq:Schrodinger}
\begin{equation}\label{eq:Schrodinger 1}
	E^{\k}_{n} \psi_{n}(\pi_{\k}) = \left\{ \frac{\pi_{\k}^{2}}{2} - \frac{k^{2}}{2} \left[ f \left( \frac{\hat{\pi}_{\k}}{M_{\star}^{1/2}} \right) \frac{\di}{\di\pi_{\k}} \right]^{2} \right\} \psi_{n}(\pi_{\k}).
\end{equation}
This can be recast as a conventional time-independent Schr\"odinger equation,
\begin{equation}\label{eq:Schrodinger 2}
	\kappa_{n} \Psi_{n}(z) = \left[ -\frac{1}{2}\frac{\di^{2}}{\di z^{2}} + V(z)\right] \Psi_{n}(z),
\end{equation}
\end{subequations}
via the  change of variables  
\begin{gather}\nonumber
	\pi_{\k} = M_{\star}^{1/2} P(z), \quad P'(z) = f(P(z)), \quad P(0) = 1, \\  \Psi_{n}(z)= \M^{{1/4}} \psi_{n}(\M^{1/2}P(z)),  
\end{gather}
and the definitions 
\be
\kappa_{n} = \frac{E^{\k}_{n}}{g^{2}\M},  \quad V(z) = \frac{P^{2}(z)}{2g^{2}}, \quad g = \frac{k}{\M}.
\ee
This shows that each deformation of the commutator maps uniquely to a potential in the Schr\"odinger
equation governing the canonical variables describing each Fourier mode. The parameter $g$ plays a central role in what follows:  large wavelength modes with $g \ll 1$ ($k\ll \M$) behave as in standard physics, but small wavelength modes with $g \gg 1$ ($k \gg \M$) exhibit exotic behaviour.  
  
\paragraph{\bf Free field propagator}

Given solutions to the eigenvalue problem (\ref{eq:Schrodinger}), it is possible to calculate the scalar field propagator. This can be accomplished with a purely quantum mechanics calculation. We begin with spatial Fourier transform of the vacuum two point function, which is given by the matrix element  
\bea
 \label{pprop-def}
D_{\k}(\tau) \equiv \langle 0_{\k}|  \hat{\phi}_{\k}(t+\tau)  
  \hat{\phi}_{\k}(t)  |0_{\k}\rangle, \quad \tau > 0.
\eea
Inserting into this the Heisenberg formula for the evolved operator $\hat{\phi}_\k(t)$ and assuming a complete energy eigenfunction basis for $\mathcal{H}_{\k}$ yields 
\begin{subequations}\label{eq:propagator}
\begin{gather}
D_{\k}(\tau) =
\sum_{n=0}^\infty e^{-i\Delta E_n^\k \tau} | c_{n}|^2 =\oint_{\mathcal C}  \frac{d\omega}{2\pi}  \; G( \omega, \k) \; e^{i\omega\tau}
\label{eq:tempD} , \\
 G( \omega, \k) \equiv  -2i \sum_{n=1}^\infty   
  \frac{  \Delta E_n^\k\  | c_n |^2} {-\omega^2 + (\Delta E_n^\k)^2 },
 \label{G}
\end{gather}
\end{subequations}
where $\Delta E_{n}^{\k}= E_{n}^{\k}-E_{0}^{\k}$, $c_{n} = c_n(\k) = \langle 0_{\k}|   \hat{\phi}_{\k} | n_\k \rangle$, and the contour $\mathcal C$ encircles all poles on the positive real axis.  This is a direct generalization of the conventional prescription for the Wightman function; other Green functions can be obtained in a similar way.

\begin{figure}
\includegraphics[width=\columnwidth]{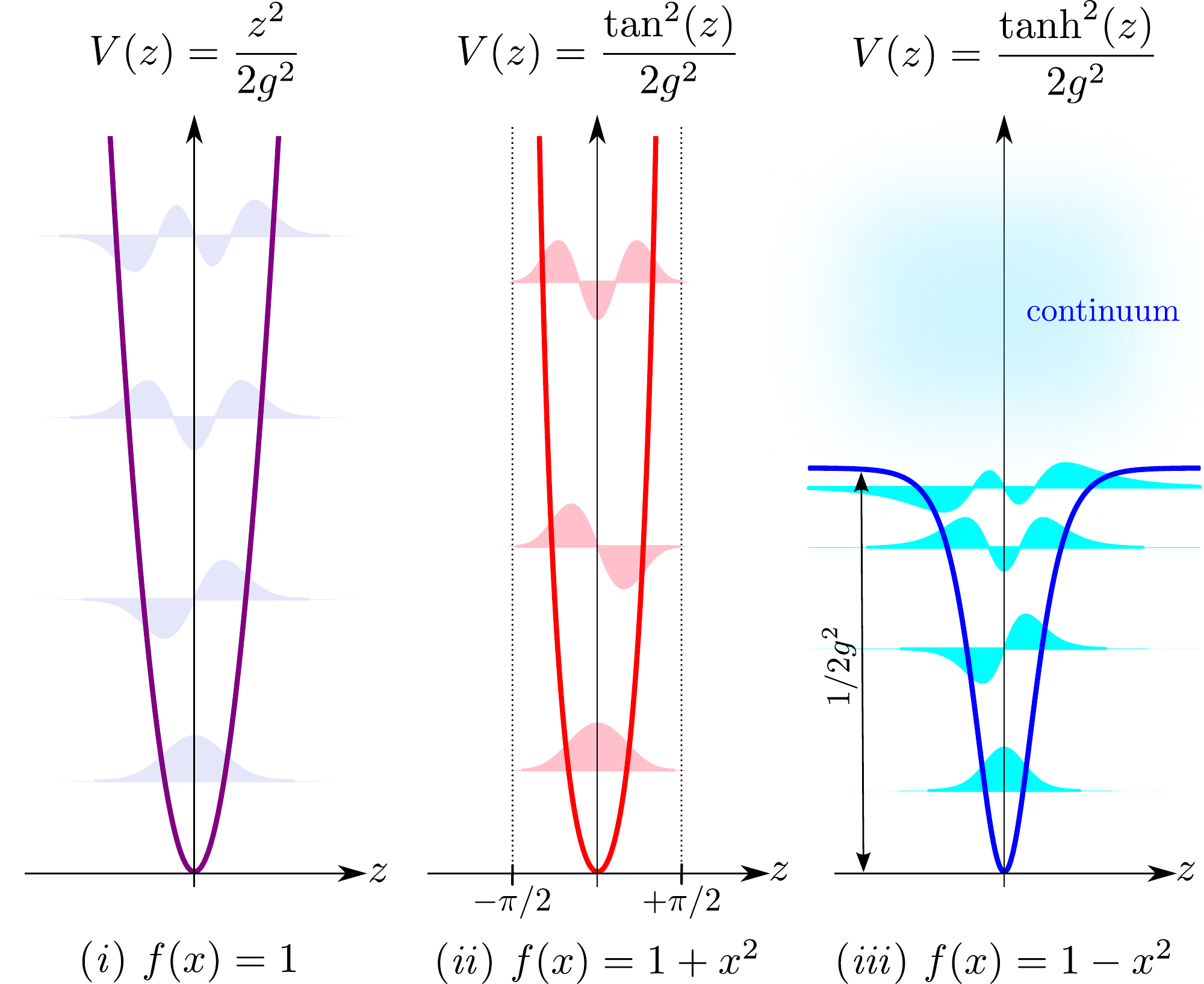}
\caption{Potentials appearing in the eigenvalue equation (\ref{eq:Schrodinger 2}) for various choices of $f$.  We take $g = 0.225$.  Note that the number of bound states in case (\emph{iii}) is directly determined by the height of the finite potential; i.e., it is a function of $g = k/\M$.}
\label{fig:potentials}
\end{figure}

The formulae (\ref{eq:propagator}) are general and apply to any quantum theory defined by the modified commutator (\ref{eq:modified commutator}). The only inputs required are the energy differences $\Delta E_n^\k$ and the matrix elements  $\langle 0_\k|   \hat{\phi}_\k | n_\k \rangle$ obtained from solutions of (\ref{eq:Schrodinger}).  It is illustrative to consider three specific choices of the function $f$ appearing in (\ref{eq:modified commutator});  the corresponding potentials appearing in the Schrodinger equation are shown in Fig. \ref{fig:potentials}.

\vspace{3mm}

\paragraph{(i) $f(x) = 1$.}

This case is the conventional quantization with $[\hat{\phi}_{\k} ,\hat{\pi}_{\k} ] = i$;  the potential in the  Schr\"odinger equation (Fig.\ \ref{fig:potentials} is that of 
the simple harmonic oscillator.  Hence,
\be
\Delta E_n^\k = nk, \quad \langle 0_{\k}|   \hat{\phi}_{\k} | n_\k \rangle  =  (2k)^{-1/2} \delta_{n,1}.
 \ee
This gives the usual answer for the Green's function 
 \be \label{eq:f=1 greens}
 G(\omega, \k) = -i/(-\omega^2 + k^2).
 \ee 
 We emphasize two key features responsible for the emergence of Lorentz invariance in the final step: (i) the exact cancellation of $k$ factors in the numerator of Eq.~(\ref{G}), and (ii) the fact that in standard quantization, $\Delta E_1^\k = k$, which gives the combination $-\omega^2 + k^2$ in the denominator.
 This provides a curious connection between the simple harmonic oscillator and Lorentz invariance, which does not exist for the  potentials associated 
 with the  deformed commutator. 
 
\vspace{3mm}

\paragraph{(ii)  $f(x) = 1 + x^{2}$.}

\begin{figure}
\includegraphics[width=\columnwidth]{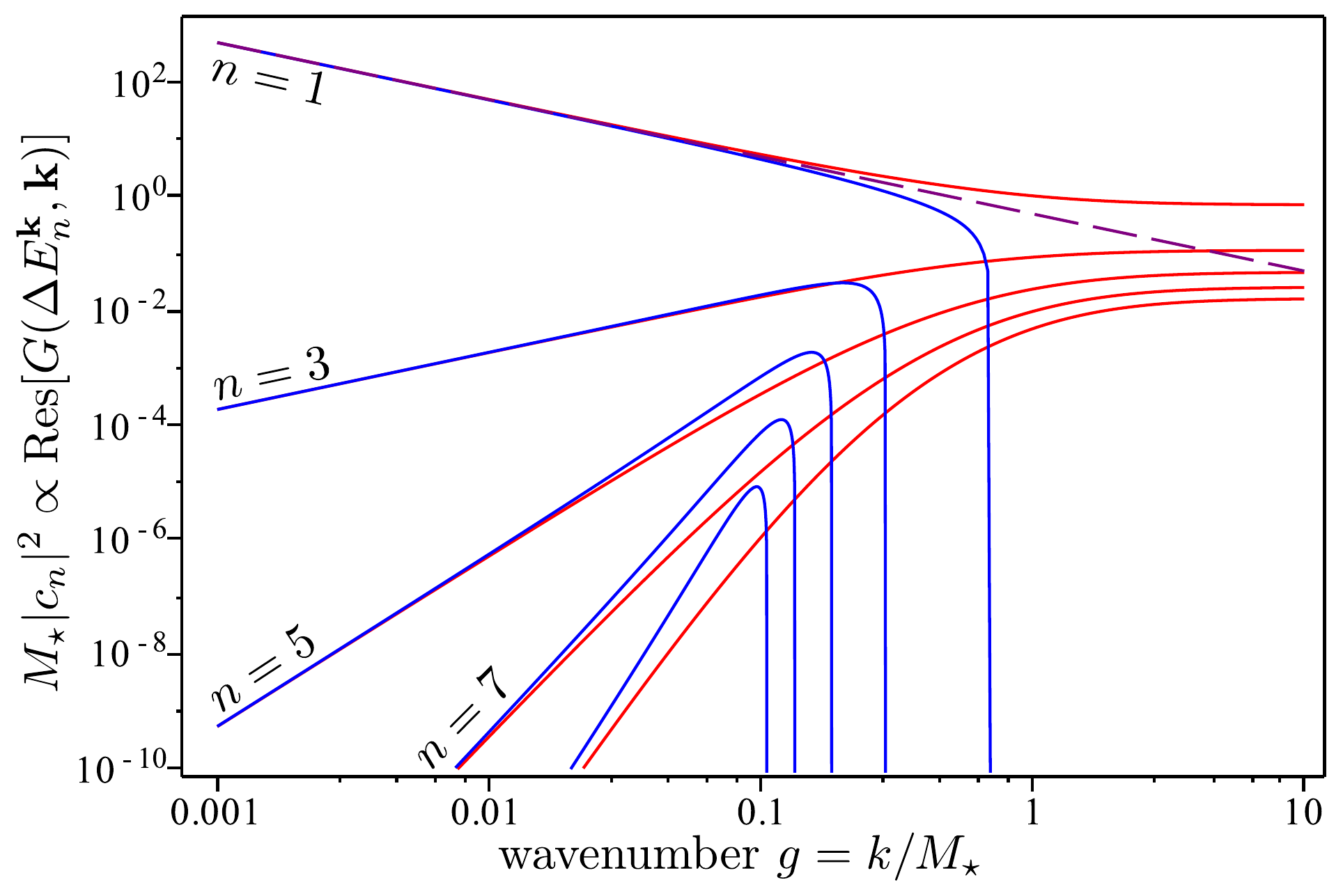}
\caption{Matrix elements $\M |c_{n}|^{2}$ for $f = 1+x^{2}$ (red), $f=1-x^{2}$ (blue), and $f = 1$ (purple dashed).  These are proportional to the residues associated with the resonant poles of the Green's function.}
\label{fig:residues}
\end{figure}

This form is motivated by including the gravitational interaction in discussions of the Heisenberg microscope gedanken experiments \cite{Mead}.   It has been widely studied at the quantum mechanics level, see eg. \cite{Das:2008}, but not in QFT.  The commutator algebra is of the form (\ref{eq:QM commutator}) with $\beta > 0$.  The potential  in the Schr\"odinger equation (\ref{eq:Schrodinger 2}) is $V(z) = \tan^{2}(z)/2g^{2}$, and is plotted in Fig.\ \ref{fig:potentials}. The change of variables introduced in obtaining (\ref{eq:Schrodinger 2}) implies that the Hilbert space for this case is  $L^2([-\pi/2,\pi/2],dz)$.  The solution of the  eigenvalue problem can be written down analytically in terms of hypergeometric functions \cite{KMM} or Gegenbauer polynomials. The energy eigenvalues yield
\begin{equation}\label{deltaE}
\frac{\Delta E^{\k}_n}{k} =  \left( \frac{g}{2} + \sqrt{1+\frac{g^2}{4}} \right)  n + 
\frac{g}{2} n^2.
\end{equation}
We have obtained analytic expressions for $c_n(\k)$, which are plotted in Fig.\ \ref{fig:residues}.  Note that $c_{2n} = 0$ for all $g$ due to parity.  The nonzero matrix elements have the following limiting behaviour
\begin{equation}
	|c_{2n-1}|^{2} \approx \frac{1}{\M} \begin{cases}
	\displaystyle  \frac{4\Gamma^{3}(n+1/2)g^{2n-3}}{\pi^{3/2} (2n-1)^{2} \Gamma(n) } , & g \ll 1, \\
	\displaystyle \frac{64 n^{2}} {\pi^{2}\left(4 n^2  - 1\right)^{2}}, & g \gg1.
	\end{cases}
\end{equation}
From this formula or Fig.\ \ref{fig:residues}, we see that for small $g$, the first matrix element coincides with the $f(x)=1$ result $|c_{1}|^{2} = (2k)^{-1}$.  The higher $n$ contributions are suppressed by successive powers of $g^{2}$, therefore the sums in (\ref{eq:propagator}) are dominated by the first term.  Since    
$\Delta E_{n}^{\k} \approx nk$ in this regime,   the propagator (\ref{eq:f=1 greens}) is recovered for $k \ll \M$.

\begin{figure}
\includegraphics[width=\columnwidth]{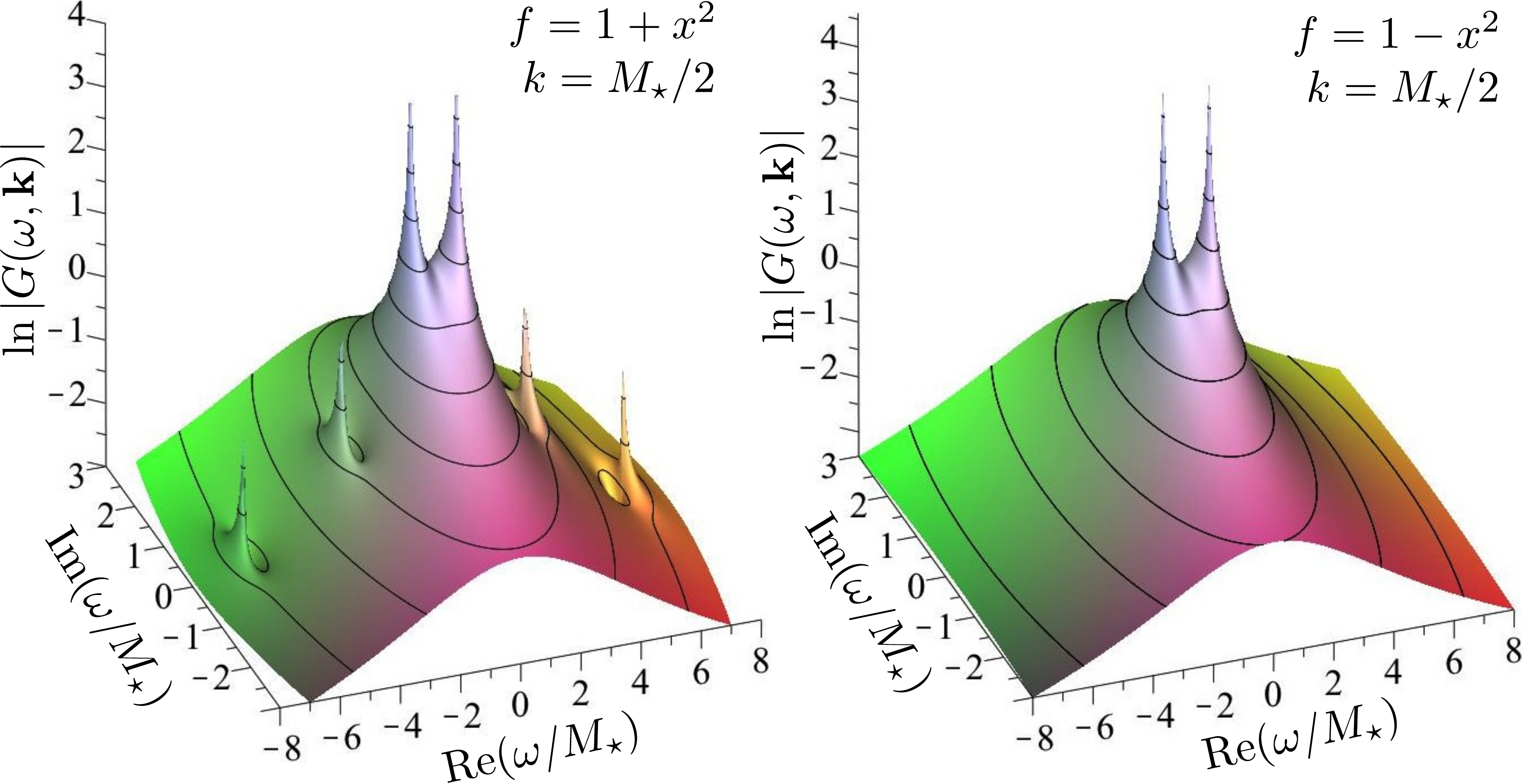}
\caption{Bound state contributions to the Green's function in the complex $\omega$ plane for $f = 1 \pm x^{2}$.  Each pair of spikes indicates a resonant mode.  The $f=1-x^{2}$ case has only two bounds states for this choice of parameters, and hence exhibits fewer poles than the $f=1+x^{2}$ case.}
\label{fig:poles}
\end{figure}

\begin{figure}
\includegraphics[width=0.8\columnwidth]{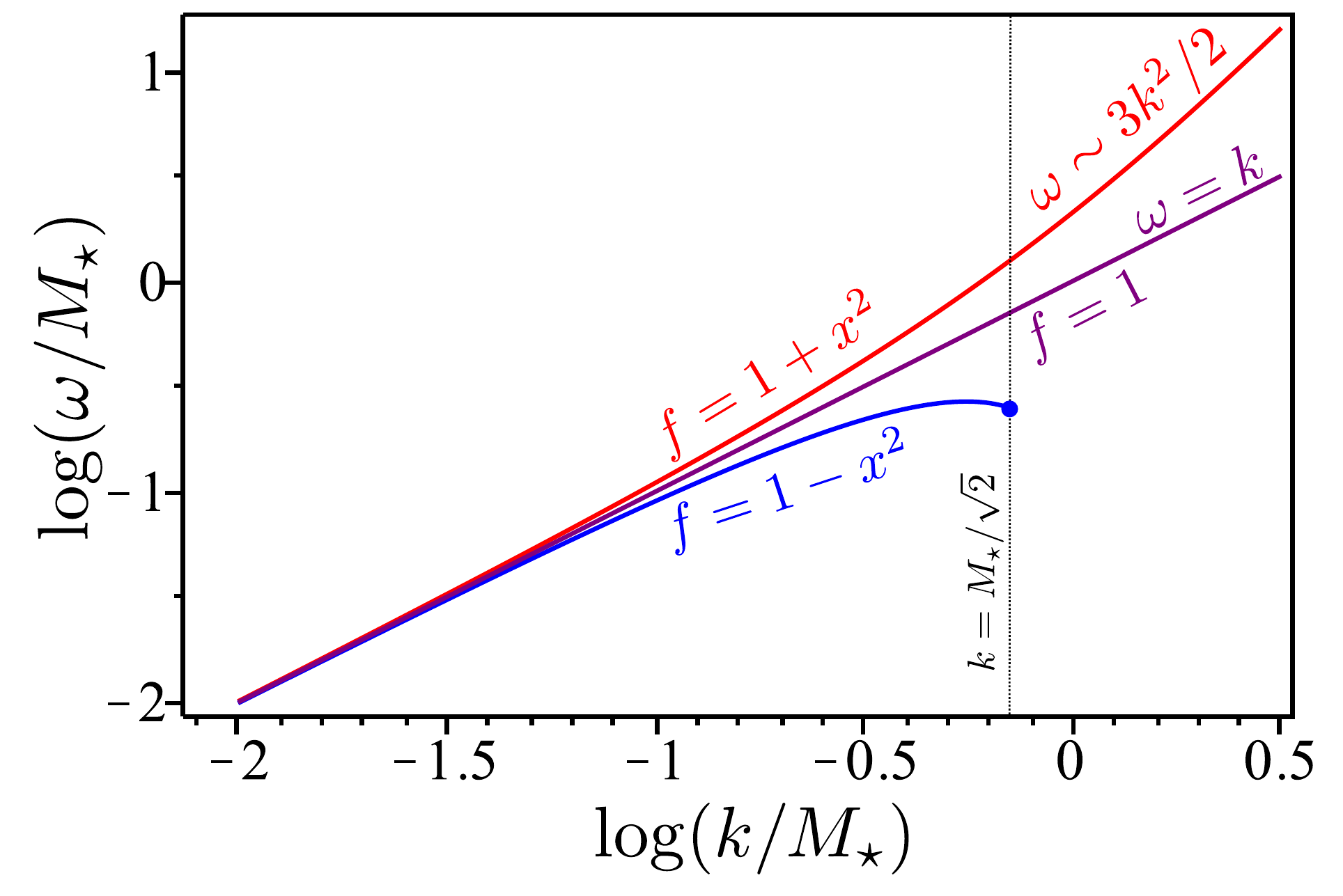}
\caption{Dispersion relations for the principal $n=1$ resonance of $G(\omega,\k$) for various choices of $f$.  Note that for the $f=1-x^{2}$ case, the $n=1$ bound state transitions to the continuum at $k = \M/\sqrt{2}$; hence the termination of the $\omega= \omega(k)$ curve.  Dispersion relations for the higher order excitations are qualitatively similar.}
\label{fig:dispersion}
\end{figure}

For $g \gtrsim 1$, the $n>1$ terms in (\ref{eq:propagator}) cannot be neglected; each of these contributes a pair of poles at $\omega = \pm \Delta E_{n}^{\k}$ to the Green's function $G(\omega,\k)$ as depicted in Fig.\ \ref{fig:poles}.  These poles may be interpreted as discrete resonant modes  with dispersion relation $\omega^{2} = (\Delta E_{n}^{\k})^{2}$.  Since the residues of the $n=1$ poles are always greater than those for  $n>1$, we call it the ``principal resonance''; its dispersion relation is shown in Fig.\ \ref{fig:dispersion}.

Finally, we note that for $g \gg 1$, the Green's function reduces to
\begin{equation}
G(\omega, \k) \approx \frac{i}{\M} \sum_{n=1}^{\infty}
\alpha_{n} \left( \frac{1}{\omega -  \frac{k^{2}}{2m^\text{eff}_{n}}} - \frac{1}{\omega +  \frac{k^{2}}{2m^\text{eff}_{n}}} \right),
\end{equation}
with $\alpha_{n} =  {64n^{2}}/{\pi^{2}(4n^{2}-1)^{2}}$ and $m_{n}^\text{eff} = {\M}/(4n^{2}-1)$.  The terms in brackets are the  Fourier-space propagators of the non-relativistic Schr\"odinger equation for a free particle of mass $m^\text{eff}_{n}$.  This result is a curious surprise: the small wavelength limit of the propagator exhibits the \emph{Galilean} symmetry  of Newtonian mechanics, and indicates  non-locality in space in this regime.  The positive and negative energy poles reflect the fact that this comes from a relativistic theory in the long wavelength limit. 

\vspace{3mm}

\paragraph{(iii)  $f(x) = 1 - x^{2}.$}

The commutator algebra in this case is of the form (\ref{eq:QM commutator}) with $\beta < 0$.  The potential appearing in the Schr\"odinger equation (\ref{eq:Schrodinger 2}) is $V(z) = \tanh^2(z)/2g^2$ which is a vertical translation of the well-known  P\"oschl-Teller potential $V(z) = -\frac{1}{2} j(j+1)\text{sech}^{2}(z)$, but  with arbitrary amplitude (see eg.\ \cite{arias-etal}).  The eigenvalue problem is again analytically solvable. However, there is a crucial difference between this case and the previous two:  the fact that  $V(z)$ has a finite height implies that for any given value of $g = k/\M$, there is a finite number of normalizable energy eigenstates in $L^{2}(\mathbb{R},dz)$. These states are labelled by integers $n=0 \ldots n_\text{max}$, where $n_{\rm max} = \text{floor} [ (\sqrt{ 4 + g^2} - g )/2g ]$.   The energy differences $\Delta E_{n}^{\k}$ for this case are given by (\ref{deltaE}) if we substitute $g \mapsto -g$.
There are no normalizable eigenstates in $L^{2}(\mathbb{R},dz)$ with energy greater than $E_\text{max}  = \M/2$. 

Because of their finite number, energy eigenfunctions do not form a complete basis of $L^{2}(\mathbb{R},dz)$ and the formulae (\ref{eq:propagator}) are not directly applicable.  However, there is a complete energy eigenfunction basis if we instead use the Hilbert space $L^{2}([-\ell,\ell],dz)$. With this  choice, the ``scattering'' energy eigenstates with $E > \M/2$ are normalizable and discrete, and the sums in (\ref{eq:propagator}) are well-defined.  Taking the $\ell \rightarrow \infty$ limit, the scattering states approach a continuum and the Green's function is
\begin{multline}\label{eq:f=1-x^2 Greens}
      G( \omega, \k) =  -2i \sum_{n=1}^{n_\text{max}}   
      \frac{ \Delta E_n^\k\  | c_n |^2} {-\omega^2 + (\Delta E_n^\k)^2 } \\ - \frac{i}{\pi} \int_{0}^{\infty} d\nu \frac{\Delta E_{\k}(\nu) |c(\nu)|^{2}}{-\omega^2 + \Delta E_{\k}^{2}(\nu) } .
\end{multline}
The integration is over the scattering states, which are labelled by the continuous parameter $\nu$.  The energy of a given scattering state is
\begin{equation}
        E_{\k}(\nu) = \frac{\M}{2}(\nu^{2}g^{2}+1), \quad \Delta E_{\k}(\nu) = E_{\k}(\nu) - E_{0}^{\k}. 
\end{equation}
Also, $c(\nu) = \lim_{\ell \rightarrow \infty} \langle \nu_{\k} | \hat\phi_{\k} |0_{\k} \rangle$ where $|\nu_{\k}\rangle$ is an odd-parity scattering mode of energy $E_{\k}(\nu)$ with normalization $\langle \nu_{\k} | \nu_{\k} \rangle = 2\ell$.  (Even parity modes do not contribute by symmetry.)

\begin{figure}
\includegraphics[width=0.9\columnwidth]{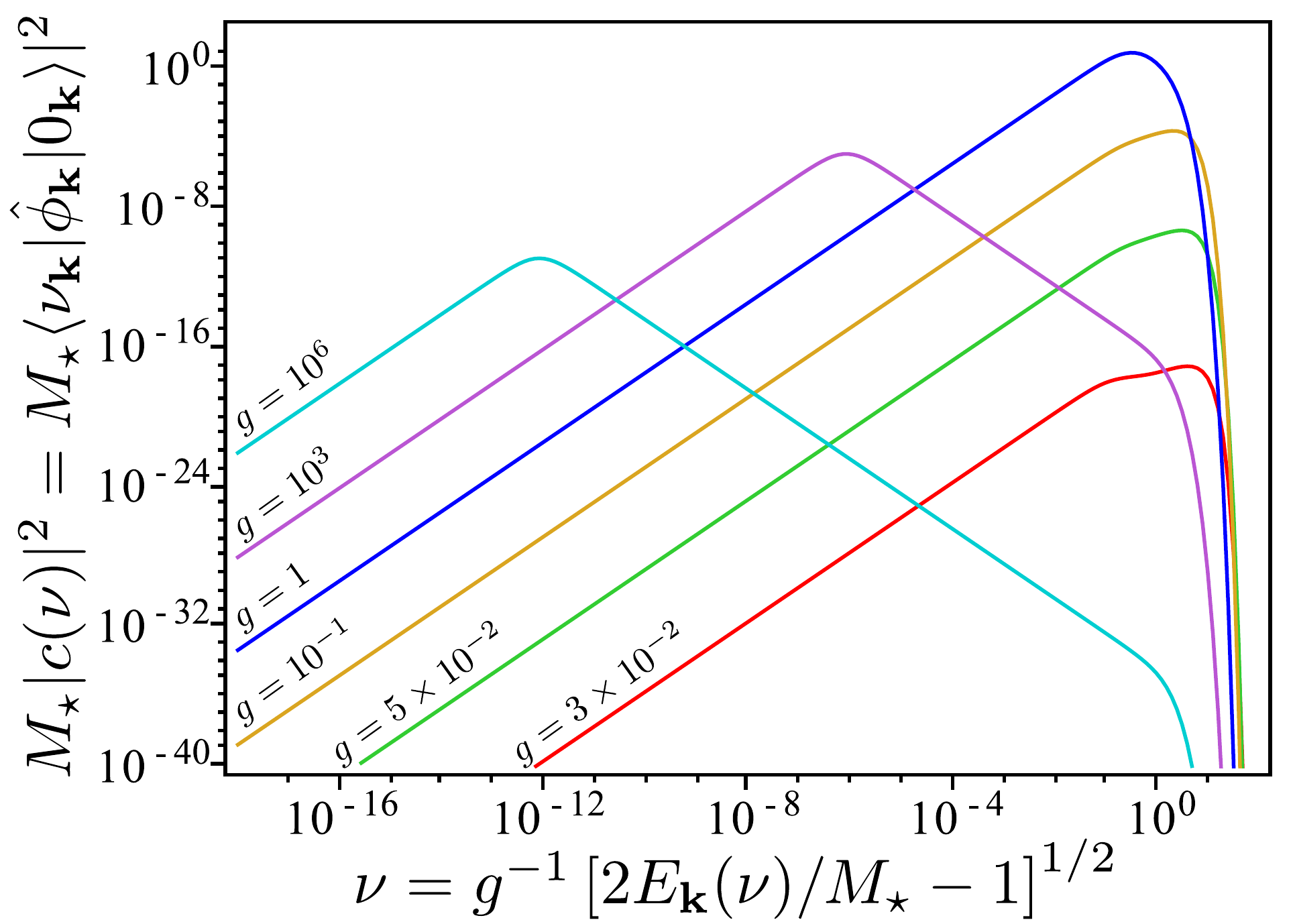}
\caption{Scattering to bound state matrix elements $|c(\nu)|^{2}$ for the $f(x)=1-x^{2}$ case.}
\label{fig:continuum}
\end{figure}
We have computed  closed form expressions for both $|c_{n}|^{2}$ (plotted in Fig.\ \ref{fig:residues}) and $|c(\nu)|^{2}$ (plotted in Fig.\ \ref{fig:continuum}).  At small $g$, the bound state residues match those for the $f(x) = 1+x^{2}$ case, as do the energy differences $\Delta E_{n}^{\k}$.  Furthermore, we find $|c(\nu)| \rightarrow 0$ as $g \rightarrow 0$, which guarantees that (\ref{eq:f=1 greens})  is recovered for long wavelengths.

Unlike the previous case, the bound state matrix elements go to zero for finite $g$.  In fact, for $g \ge 1/\sqrt{n(n+1)}$ we have $c_{n}=0$.  The vanishing of $c_{n}$ at particular values of $g = k/\M$ indicates the threshold where the $n^\text{th}$ eigenstate swtiches from a bound to scattering state, or vice versa.  It follows that for $g \ge 1/\sqrt{2}$, the only non-zero contribution to Green's function (\ref{eq:f=1-x^2 Greens}) is from the continuum integral.  For $g\gg 1$, it can be shown that
\begin{equation}
	D_{\k}(\tau) \approx \frac{8  \sqrt{2 \pi} e^{-i(\M^{3}\tau/2k^{2}+3\pi/4)}}{k(\M\tau)^{3/2}} F \left( \frac{\tau\M^{3}}{k^{2}} \right),
\end{equation}
where $F(u)=1 + \mathcal{O}(1/u)$ is a non-oscillatory  expression involving error functions.  The $\tau^{3/2}$ factor in the denominator implies that short wavelength modes decay to zero with characteristic timescale $\M^{-1}$, and  the argument of the exponential gives an effective dispersion relation 
$\omega = \M^{3}/2k^{2}$.  That is, disturbances of physical size $\ll \M^{-1}$ will not generate any long distance wave propagation in this model. It is interesting
that some of the features of the $f(x)=1-x^2$ case appear  to naturally incorporate  recent ideas on the ultraviolet completion of  non-renormalizable theories via a so-called ``classicalization''  \cite{Dvali:selfc},  where propagating quantum degrees of freedom do not exist at short distance scales.

\vspace{3mm}

\paragraph*{\bf Discussion}
 We have shown that if modified canonical commutators are directly  implemented in  $\k$ space,  the resulting QFTs  exhibit novel short
wavelength behaviour: propagation amplitudes have multiple poles,  Lorentz invariance is broken,  and  there is spatial non-locality.

The expression for  the  propagator   Eq.~(\ref{G}) resembles the Lehmann-Kallen spectral representation for an interacting Lorentz invariant scalar
field theory of mass $m$, 
 \be
-i G(\omega, \k) = \frac{1}{-\omega^2 +\k^2 + m^2} + \int \limits_{4m^2}^{\infty} \frac{d \mu^2 \; \rho(\mu^2)}{-\omega^2 +\k^2 + \mu^2},
 \ee
where $\rho(\mu^2)$ is the density of particle resonances of mass $\mu$. Although purely mathematical, this analogy 
suggests a multi-particle nature of deformed quantization, even for a free theory. (Our result on multiple poles  support an argument to
that effect in  \cite{Hossenfelder:2007} for QFTs with a minimal length scale based on higher derivative theories.) 
 
 Another interesting outcome of this work is an explicit non-locality in space at short distance scales.  This  may be seen in at least two distinct ways.  One is simply due to Lorentz violation; in the large $k$ regime for the case  $f=1+x^2$, the non-localilty is exactly Newtonian action-at a-distance, as evident from the form of the propagator.  The other follows directly from inverse Fourier transformation of the deformed $\k$ space commutator, which would lead to non-local
terms of the form $M^{-1}_{\star} \int d^3 \mathbf{z} \, \hat\pi(t,\mathbf{y}-\mathbf{z}) \, \hat \pi(t,\mathbf{z})$
in the equal time commutator $[ \hat \phi(t,\mathbf{x}), \hat \pi(t,\mathbf{x}+\mathbf{y}) ]$.
The Heisenberg equations of motion derived from this commutator would be non-local integro-differential equations. 
 
The above observation appears to  resonate with the recently proposed hypothesis of relative non-locality \cite{rel-loc}, in which locality in a postulated
curved momentum space leads to non-locality in physical space.   In our approach, space non-locality is evident in the above commutator, and the   
``relative" part may be connected with the fact that our quantization depends on  a choice of  time coordinate. Since that hypothesis   is
motivated by earlier works on deformed Lorentz symmetry, perhaps such a connection is not unexpected, with the caveat that in our approach it
arises at the level of a dynamical quantum field,  rather than at the kinematical level  in \cite{rel-loc}.  

There are several directions for further work using the deformed quantization we have discussed. It is apparent that the approach may be 
 followed for  other spin fields in flat space time.  Of particular interest is the scalar field on a  black hole background; for example, modified dispersion relations
of the type computed for the $f_{-}$ case, with vanishing group velocity for short wavelength modes,  have been used to study Hawking radiation
\cite{unruh-dispersion}. 

For interacting theories, perhaps of most interest for  observational consequences is quantum electrodynamics.  Of related 
interest is whether Lorentz violation from this type of deformed quantization gives rise to order one effects due to loop corrections,
 as discussed in the context of effective field theory (EFT) in \cite{collinsLV}. However, many components of  conventional
 quantization are \emph{prima-facie} absent in our deformed quantization, so it is not clear whether the axioms of EFT apply.  Indeed, space non-locality at short
 wavelengths appears to complicate the Wilsonian program of integrating out  high energy degrees of freedom and capturing their effects in
 local counter-term; related comments on this appear in \cite{Dvali:selfc}. Thus, any arguments on Lorentz violation based on EFT would have to
 be carefully reformulated  before making statements concerning renormalization and loop corrections.

\smallskip

This work was supported by an  Atlantic Association for Research in the Mathematical Sciences (AARMS) post-doctoral fellowship (D.K.) and
NSERC of Canada.

\bibliographystyle{apsrev}
\bibliography{gup-refs}

\end{document}